\documentclass[aps,prd,reprint,superscriptaddress]{revtex4-2}
\usepackage{amsmath,amssymb,multirow,bm}
\usepackage{soul}
\usepackage{epsfig}
\usepackage{graphicx}
\usepackage{lipsum}
\usepackage{orcidlink}
\usepackage{color}
\makeatother

\usepackage{babel}
\usepackage{hyperref}
\preprint{}

\begin{document}
\title{Neutron stars in a conservative $f(R,T)$ gravity}

\author{Ronaldo V. \surname{Lobato} \orcidlink{0000-0001-5755-5363},}
\affiliation{Centro Brasileiro de Pesquisas F\'isicas, Rua Dr. Xavier Sigaud 150, Rio de Janeiro 22290-180, RJ, Brazil}

\author{Geanderson A. \surname{Carvalho} \orcidlink{0000-0001-8718-6925},}
\affiliation{Departamento de F\'isica, Universidade Tecnol\'ogica Federal do Paran\'a, Medianeira, PR, Brazil}
\affiliation{Programa de P\'os-Gradua\c{c}\~ao em F\'isica e Astronomia,
Universidade Tecnol\'ogica Federal do Paran\'a, Jardim das Am\'ericas 82590-300, Curitiba, PR, Brazil}
\affiliation{N\'ucleo de Astrof\'isica e Cosmologia (Cosmo-Ufes), Universidade Federal do Esp\'irito Santo, Av. Fernando Ferrari 514, Vit\'oria, 29075-910, ES, Brazil}

\author{Carlos E. C. \surname{Monta\~na} \orcidlink{0000-0002-6442-4668},}
\affiliation{Facultad de Ciencias Básicas, Universidad Santiago de Cali, Calle 5 62-00, Pampalinda, Cali, Valle del Cauca, Colombia}

\author{Jose F. \surname{Rodriguez-Ruiz} \orcidlink{0000-0003-3627-5084},}
\affiliation{Departamento de F\'isica, Universidad Antonio Nari\~no, Cra 3 Este \# 47A - 15, Bogot\'a D.C. 110231, Colombia}

\date{\today}

\begin{abstract}
We investigate a conservative formulation of $f(R,T)$ gravity motivated by a key limitation of several existing approaches: the gravitational function is often reconstructed from a chosen equation of state, making the gravity sector EoS-dependent and compromising universality. To avoid this problem, we reformulate the theory in terms of an effective energy-momentum tensor, so that the conservation law follows from the field equations and Bianchi identities while the gravitational action remains independent of the microphysical EoS. We derive the modified stellar structure equations, establish theoretical consistency conditions including coupling bounds and crust-singularity avoidance, and present the tidal perturbation sector in terms of effective thermodynamic variables and an effective sound speed. We then compute neutron star observables using realistic tabulated EoSs, including mass-radius relations and tidal deformabilities, and compare the model with current astrophysical constraints from massive pulsars, NICER radius measurements, and GW170817.
\end{abstract}

\maketitle

\section{Introduction}

Understanding the behavior of matter and gravity under extreme
conditions is a central goal of modern astrophysics. Compact
objects such as neutron stars provide a unique laboratory for
probing the properties of dense matter and testing gravitational
theories in the strong-field regime. Observations of massive
pulsars with masses close to two solar masses
\cite{demorest/2010, antoniadis/2013, cromartie/2019, fonseca/2021},
together with radius measurements from the NICER mission
\cite{riley/2019, miller/2019, riley/2021, miller/2021} and tidal deformability constraints from gravitational wave observations of
binary neutron star mergers \cite{abbott/2017a}, have
significantly improved our ability to test both the microphysics
of dense matter and the underlying theory of gravity.

General Relativity has been remarkably successful in describing
gravitational phenomena over a wide range of scales.
Nevertheless, theoretical and observational motivations continue
to drive the exploration of modified theories of gravity
\cite{sotiriou/2010, defelice/2010, nojiri/2011}. Neutron stars are
particularly powerful probes of such modifications in the strong-field regime \cite{yazadjiev/2015, staykov/2014a,
capozziello/2016, astashenok/2021, astashenok/2022}. Among these
alternatives, $f(R)$ gravity and its extensions have attracted
considerable attention. A particularly interesting
generalization is the $f(R,T)$ theory of gravity, originally
proposed by Harko \textit{et al.} \cite{harko/2011}, in which the
gravitational action depends not only on the Ricci scalar $R$
but also on the trace $T$ of the energy-momentum tensor. The
trace dependent terms can be interpreted as an effective
coupling between matter and geometry, which may arise from
quantum effects \cite{lobato/2019} or from the presence of imperfect fluids \cite{yousaf/2016}.

A generic feature of $f(R,T)$ gravity is that the covariant
divergence of the energy-momentum tensor does not vanish in
general, $\nabla_\mu T^{\mu\nu} \neq 0$
\cite{harko/2011, barrientoso./2014}, reflecting the nonminimal
coupling between matter and curvature. This feature has
motivated the development of conservative formulations in which
one instead imposes $\nabla_\mu T^{\mu\nu}=0$. In particular, it
has been proposed that this condition can be used to determine
the functional form of $h(T)$ in models of the type
$f(R,T)=R+h(T)$ \cite{chakraborty/2013}.
Such formulations have been explored in both cosmological and
astrophysical settings, including neutron stars
\cite{dossantos/2019, pretel/2021a} and strange quark stars
\cite{carvalho/2020}.

A conceptual difficulty with several energy-momentum conserving
formulations of $f(R,T)$ gravity is that the functional form of
the gravitational action is obtained by solving the conservation
condition for a specific equation of state. In these approaches,
the function $h(T)$ in models of the form $f(R,T)=R+h(T)$ is
determined by combining the conservation equation with a chosen
relation between pressure and energy density, such as a
barotropic equation of state $p=\omega\rho$, a polytropic model
$p=K\rho^{\Gamma}$, or the MIT bag model used to describe
strange quark matter. The same conceptual issue arises for all
these parameterizations. In barotropic models, changing
$\omega$ changes the stiffness and therefore the reconstructed
$h(T)$; in polytropic models, different choices of $(K,\Gamma)$
play the same role; and in MIT bag descriptions, varying $B$
(and $\omega$, when treated as free) likewise changes the
inferred $h(T)$. Thus, even when these EoSs are used as
effective representations of uncertain microphysics, the
reconstructed gravity function tracks modeling choices rather
than a unique underlying theory.

In this situation, the gravitational Lagrangian becomes
dependent on the microphysical properties of matter,
effectively reversing the usual hierarchy in which the
gravitational theory should be specified independently of the
material sector. From this perspective, the result is not a
unique modification of gravity but rather a family of theories
associated with different equations of state.

This issue becomes particularly relevant in the modeling of
neutron stars. Realistic nuclear equations of state encode a
substantial amount of microphysical information about dense
matter \cite{oertel/2017a}, including nuclear many-body
interactions, three-body forces, the density dependence of the
symmetry energy, and possible phase transitions to hyperonic or
quark matter. Moreover, physically viable equations of state
must satisfy theoretical constraints such as causality. In
particular, the sound speed must obey $c_s^2 = dp/d\rho \leq 1$,
ensuring that perturbations propagate at subluminal speeds
\cite{bedaque/2015, altiparmak/2022}.

Modern neutron star equations of state are therefore obtained
from microscopic nuclear calculations and are often implemented
in tabulated form. If the gravitational action were determined
from the equation of state, each nuclear model would correspond
to a different gravitational theory. In such a scenario,
modifying the nuclear interaction model or the high-density
behavior of the equation of state would effectively modify the
underlying theory of gravity. This creates a degeneracy between
gravitational effects and uncertainties in the nuclear
microphysics (e.g., \cite{he/2015}), making it difficult to
disentangle genuine modifications of gravity from variations in
the equation of state. Similar degeneracies have been discussed
in the context of scalar-tensor and other modified gravity
theories \cite{minamitsuji/2016, yagi/2017, doneva/2023,
silva/2024}.

A further conceptual difficulty concerns the universality of
the gravitational interaction. In metric theories of gravity,
the form of the gravitational action is universal and does not
depend on the particular properties of the matter configuration
under consideration. If the functional form of $f(R,T)$ depends
on the equation of state, then the gravitational theory would
effectively change when different astrophysical systems are
studied. For example, neutron stars described by nuclear
equations of state, strange quark stars described by the MIT
bag model, and cosmological fluids characterized by different
equations of state would correspond to different gravitational
Lagrangians.

This lack of universality is closely related to the Einstein
equivalence principle, which states that the outcome of local
non-gravitational experiments is independent of the composition
and internal structure of matter. If the gravitational dynamics
depend explicitly on the equation of state, then the
gravitational interaction would effectively depend on the
microscopic properties of the fluid describing the system,
blurring the distinction between modifications of gravity and
changes in the microphysical description of matter.

Motivated by these considerations, we investigate a formulation
of $f(R,T)$ gravity in which the gravitational action is
specified independently of the equation of state and the trace dependent contributions are interpreted in terms of an
effective energy-momentum tensor. In this approach, the
conservation law emerges naturally from the field equations
through the Bianchi identities, without imposing additional
constraints that relate the gravitational Lagrangian to the
microphysical properties of matter.

We apply this formulation to compact stars and derive the
corresponding stellar structure equations. Using realistic
neutron star equations of state obtained from nuclear many-body
calculations, we investigate the impact of the modified gravity
terms on the mass-radius relation and on the tidal properties
of neutron stars. In particular, tidal deformability, which
plays a central role in the gravitational-wave signal from
binary neutron star mergers, provides an important
observational probe of deviations from General Relativity. By
solving the stellar structure equations together with the
differential equation governing the tidal Love number, we
compute the mass-radius relations and tidal deformabilities
predicted by the model and compare them with current
astrophysical constraints from massive pulsars
\cite{demorest/2010, antoniadis/2013}, NICER measurements
\cite{riley/2019, miller/2019}, and gravitational-wave
observations such as GW170817.

\section{Conservative $f(R,T)$ Models}

Several works in the literature attempt to construct energy-momentum
conserved versions of $f(R,T)$ gravity by imposing the condition
\begin{equation}
\nabla_\mu T^{\mu\nu}=0
\end{equation}
directly on the field equations in order to determine the functional
form of the trace dependent function $h(T)$ in models of the form
\begin{equation}
f(R,T)=R+h(T).
\end{equation}

In these approaches the conservation condition is solved simultaneously
with a specific equation of state (EoS) describing the matter sector.
For example, assuming a barotropic equation of state
\begin{equation}
p=\omega\rho
\end{equation}
leads to a differential equation for $h_T = dh/dT$. The conservation equation is
\begin{equation}
\label{eq:conserv}
(\rho+p)\frac{d}{dr}\ln h_T+\frac{1}{2}\frac{d}{dr}(\rho+3p)=0,
\end{equation}
so
\begin{equation}
\frac{d\ln h_T}{d\rho}=-\frac{1+3\omega}{2(1+\omega)}\frac{1}{\rho}
\quad\Rightarrow\quad
h_T(\rho)\propto \rho^{-\frac{1+3\omega}{2(1+\omega)}}.
\end{equation}
Using $T=\rho-3p=(1-3\omega)\rho$ (for $\omega\neq 1/3$), one has
\begin{equation}
h_T(T)=\frac{dh}{dT}\propto T^{-\frac{1+3\omega}{2(1+\omega)}},
\end{equation}

\begin{equation}
h(T) \propto T^{\frac{1-\omega}{2(1+\omega)}} .
\end{equation}
where $\omega\neq -1$ is also required.

If a different equation of state is adopted, such as the MIT bag model
for strange quark matter,
\begin{equation}
p=\omega(\rho-4B),
\end{equation}
a different functional form of $h(T)$ is obtained.

For the realistic nuclear EoSs, the conservation condition, Eq.~\eqref{eq:conserv}, fixes $h_T(\rho)$ and therefore $h(T)$. In practice, the reconstructed
function depends on the chosen matter model, i.e. $h(T)\equiv h(\rho,p(\rho))$,
so the action is effectively
\begin{equation}
f(R,T)=R+h(\rho,p(\rho)).
\end{equation}
Thus, the gravity sector becomes explicitly EoS-dependent. If one uses a database such as COMPOSE \cite{typel/2015}, which contains hundreds of models, one effectively obtains hundreds of gravity models.

This introduces a structural degeneracy in compact-star modeling: both
EoS stiffness and the reconstructed gravity correction modify the same
observables (mass-radius curves and tidal deformabilities). Since $h(T)$
is built from the EoS itself, part of what appears as a gravity signal can
be reabsorbed into matter modeling, and vice versa. This leads to the following:

\paragraph{Non-uniqueness of the gravitational action}

For the class $f(R,T)=R+h(T)$, imposing $\nabla_\mu T^{\mu\nu}=0$ after
choosing an EoS does not compare different matter sectors within one fixed
gravity theory. Instead, each EoS choice (barotropic, polytropic, MIT bag,
or realistic tabulated nuclear EoS) generates a different reconstructed
$h(T)$ and therefore a different effective Lagrangian.

\paragraph{Implications for neutron star inference}

In neutron star applications this mapping is particularly limiting, because
realistic EoSs form families constrained by nuclear theory and observations
rather than a unique input. As a consequence, varying EoS stiffness and
varying gravity are not independent operations, which weakens direct
interpretability of gravity constraints extracted from astrophysical data.

\paragraph{Universality and equivalence-principle motivation}

A fundamental gravity model should be defined once and then applied across
systems (neutron stars, strange stars, cosmology) without changing the
action. If $h(T)$ is reconstructed separately for each EoS class, this
universality is blurred, and the effective dynamics acquires composition-
dependent modeling choices from the matter sector.

These issues motivate the strategy adopted here: we specify the
gravitational action a priori and absorb the trace dependent terms into an
effective energy-momentum tensor,
\begin{equation}
G_{\mu\nu}=8\pi T^{\rm eff}_{\mu\nu}.
\end{equation}
Then, from the Bianchi identity
\begin{equation}
\nabla_\mu G^{\mu\nu}=0,
\end{equation}
we obtain automatically
\begin{equation}
\nabla_\mu T^{\rm eff\,\mu\nu}=0.
\end{equation}
In this formulation the EoS enters only at the level of matter closure in
stellar-structure calculations, not in the definition of the gravity theory.
This preserves a single gravitational framework while keeping the gravity-EoS degeneracy explicit and testable.

\section{An effective energy-momentum tensor conservative $f(R,T)$ model}
\label{sec:model}
\subsection{Action and Field Equations}

We consider the model $f(R,T) = R + 2\chi T$ \cite{harko/2011}, where $R$ is the Ricci scalar, $T$ is the trace of the energy-momentum tensor, and $\chi$ is a constant coupling parameter. For this model, the gravitational action is
\begin{equation}
S=\int\left[\frac{R+2\chi T}{16\pi}+L_m\right]\sqrt{-g}\,d^4x ,
\end{equation}
where $L_m$ is the matter Lagrangian density.

Assuming the standard perfect-fluid choice $L_m=-p$ and varying with respect to the metric yields
\begin{equation}
G_{\mu\nu}=
8\pi T_{\mu\nu}
+2\chi(T_{\mu\nu}+p g_{\mu\nu})
+\chi T g_{\mu\nu}.
\end{equation}

For the perfect fluid, with $u^\mu u_\mu=1$,
\begin{equation}
\label{em:tensor}
T_{\mu\nu}=(\rho+p)u_\mu u_\nu-pg_{\mu\nu},
\end{equation}

and

\begin{equation}
T=\rho-3p.
\end{equation}

\subsection{Effective Energy-Momentum Tensor}

The field equations can be rewritten as
\begin{equation}
\label{gr:effective}
G_{\mu\nu}=8\pi T_{\mu\nu}^{\mathrm{eff}},
\end{equation}
where the effective tensor has the perfect–fluid form

\begin{equation}
T^{\mathrm{eff}}_{\mu\nu}
=(\rho_{\mathrm{eff}}+p_{\mathrm{eff}})u_\mu u_\nu
-
p_{\mathrm{eff}}g_{\mu\nu}.
\end{equation}

Matching terms gives

\begin{equation}
\label{eq:rhoeff}
\rho_{\mathrm{eff}}=\rho+\frac{\chi}{8\pi}(3\rho-p),
\end{equation}

\begin{equation}
\label{eq:peff}
p_{\mathrm{eff}}=p+\frac{\chi}{8\pi}(3p-\rho).
\end{equation}

The linear map between $(\rho,p)$ and $(\rho_{\mathrm{eff}},p_{\mathrm{eff}})$ is invertible for
$\chi\neq-4\pi$ and $\chi\neq-2\pi$. Explicitly,
\begin{equation}
\rho+p=\frac{\rho_{\mathrm{eff}}+p_{\mathrm{eff}}}{1+\chi/(4\pi)}.
\end{equation}
For practical stellar-structure integration with a physical EoS $p=p(\rho)$,
it is useful to write the inverse map explicitly:
\begin{equation}
\rho=\frac{\left(1+\frac{3\chi}{8\pi}\right)\rho_{\mathrm{eff}}+\frac{\chi}{8\pi}p_{\mathrm{eff}}}{\left(1+\frac{\chi}{4\pi}\right)\left(1+\frac{\chi}{2\pi}\right)},
\end{equation}
\begin{equation}
p=\frac{\frac{\chi}{8\pi}\rho_{\mathrm{eff}}+\left(1+\frac{3\chi}{8\pi}\right)p_{\mathrm{eff}}}{\left(1+\frac{\chi}{4\pi}\right)\left(1+\frac{\chi}{2\pi}\right)}.
\end{equation}
These relations ensure a one-to-one closure between the physical EoS and the
effective-fluid TOV system whenever the invertibility conditions above are satisfied.

Because the Einstein tensor satisfies the Bianchi identities,
\begin{equation}
\nabla_\mu G^{\mu\nu}=0,
\end{equation}
the effective tensor is conserved (whereas $T_{\mu\nu}$ is not conserved in general in $f(R,T)$ theories)
\begin{equation}
\nabla_\mu T^{\mu\nu}_{\mathrm{eff}}=0.
\end{equation}

The theory can be mathematically rewritten as General Relativity sourced by an
effective fluid, although the effective thermodynamic variables originate from
the matter-geometry coupling and are not independent microphysical fluid
variables.

\subsection{Decomposition and Effective Coupling}

The field equations can be decomposed as

\begin{equation}
G_{\mu\nu}
=(8\pi+2\chi)T_{\mu\nu}+(2\chi p+\chi T)g_{\mu\nu}.
\end{equation}

Hence, the coefficient of $T_{\mu\nu}$ may be written as

\begin{equation}
8\pi+2\chi = 8\pi\left(1+\frac{\chi}{4\pi}\right).
\end{equation}

One may define

\begin{equation}
G_{\mathrm{eff}} = G\left(1+\frac{\chi}{4\pi}\right),
\end{equation}

but this should be interpreted with care: the extra term
$(2\chi p+\chi T)g_{\mu\nu}$ does not vanish in general. Therefore, the model
is not equivalent to simply replacing $G$ by $G_{\mathrm{eff}}$ in GR; rather, it is
GR sourced by an effective fluid, as written in Eq.~\eqref{gr:effective}.

\subsection{Modified Tolman-Oppenheimer-Volkoff Equations}

We consider the static spherically symmetric metric (signature $+,-,-,-$)
\begin{equation}
ds^2=e^{\nu(r)}dt^2-e^{\lambda(r)}dr^2-r^2 d\Omega^2.
\end{equation}

Defining the mass function by
\begin{equation}
e^{-\lambda(r)}=1-\frac{2m(r)}{r},
\end{equation}
the $tt$ component of the Einstein equations gives
\begin{equation}
\frac{dm}{dr}=4\pi r^2 \rho_{\mathrm{eff}}.
\end{equation}

The $rr$ component yields
\begin{equation}
\nu'=\frac{2(m+4\pi r^3 p_{\mathrm{eff}})}{r(r-2m)}.
\end{equation}

From the conservation equation $\nabla_\mu T^{\mu\nu}_{\mathrm{eff}}=0$,
\begin{equation}
\frac{dp_{\mathrm{eff}}}{dr}
=
-\frac{\nu'}{2}(\rho_{\mathrm{eff}}+p_{\mathrm{eff}}).
\end{equation}
Substituting $\nu'$ we obtain
\begin{equation}
\frac{dp_{\mathrm{eff}}}{dr}
=
-(\rho_{\mathrm{eff}}+p_{\mathrm{eff}})
\frac{m+4\pi r^3 p_{\mathrm{eff}}}{r(r-2m)}.
\end{equation}

Therefore, the effective-fluid stellar-structure system is
\begin{equation}
\frac{dm}{dr}=4\pi r^2 \rho_{\mathrm{eff}},
\end{equation}
\begin{equation}
\frac{dp_{\mathrm{eff}}}{dr}
=
-(\rho_{\mathrm{eff}}+p_{\mathrm{eff}})
\frac{m+4\pi r^3 p_{\mathrm{eff}}}{r(r-2m)}.
\end{equation}
The standard Tolman-Oppenheimer-Volkoff equations \cite{tolman/1939, oppenheimer/1939} are recovered in the limit
\begin{equation}
\chi \rightarrow 0.
\end{equation}

\paragraph{Bridge to the physical pressure gradient:}
To connect the effective-fluid system to a physical EoS $p=p(\rho)$, we use
\begin{equation}
p_{\mathrm{eff}}=p+\frac{\chi}{8\pi}(3p-\rho),
\end{equation}
which implies
\begin{equation}
\frac{dp_{\mathrm{eff}}}{dr}=\left(1+\frac{3\chi}{8\pi}\right)\frac{dp}{dr}-\frac{\chi}{8\pi}\frac{d\rho}{dr}.
\end{equation}
Hence
\begin{equation}
\frac{dp}{dr}=\frac{1}{1+\frac{3\chi}{8\pi}}\left[\frac{dp_{\mathrm{eff}}}{dr}+\frac{\chi}{8\pi}\frac{d\rho}{dr}\right].
\end{equation}
This equation has the implicit requirement $1+3\chi/8\pi\neq 0$, i.e. $\chi\neq-8\pi/3$. This value is already excluded by the dominant-energy-condition branch used for ordinary neutron star matter ($\rho>p$), which gives $\chi\ge -2\pi$, as shown below.

For a barotropic EoS $\rho=\rho(p)$, with $c_s^2\equiv dp/d\rho$ it becomes,
\begin{equation}
\label{dpdr}
\frac{dp}{dr}=\frac{\frac{dp_{\mathrm{eff}}}{dr}}{\left(1+\frac{3\chi}{8\pi}\right)-\frac{\chi}{8\pi}\frac{1}{c_s^2}}.
\end{equation}

To avoid a coordinate singularity in the hydrostatic equation, the denominator
of Eq. \eqref{dpdr} must not vanish anywhere inside the star. Therefore,
\begin{equation}
\left(1+\frac{3\chi}{8\pi}\right)-\frac{\chi}{8\pi}\frac{1}{c_s^2} \neq 0,
\end{equation}
which implies the theoretical constraint
\begin{equation}
\chi \neq -\frac{8\pi}{3-1/c_s^2}.
\end{equation}
For causal matter ($0<c_s^2\le 1$), one also typically requires this factor to
remain positive throughout the stellar interior.

\subsection{Constraints on the Coupling Parameter}
\subsubsection{Energy Conditions}

Physical solutions require that the effective fluid satisfies

\begin{equation}
\rho_{\mathrm{eff}}>0,
\end{equation}

\begin{equation}
\rho_{\mathrm{eff}}+p_{\mathrm{eff}}>0.
\end{equation}

Using the definitions above gives the conditions

\begin{equation}
\rho+\frac{\chi}{8\pi}(3\rho-p)>0,
\end{equation}

\begin{equation}
\rho+p+\frac{\chi}{4\pi}(\rho+p)>0.
\end{equation}

The allowed parameter space of the coupling $\chi$ is constrained by three logically distinct requirements: (i) intrinsic theoretical consistency of the effective-fluid formulation, (ii) regularity of the modified hydrostatic equilibrium equations, and (iii) compatibility with astrophysical observations.

\subsubsection*{Theoretical Constraints}
Under the standard neutron star matter assumptions
\begin{equation}
\rho+p>0,\qquad \rho>p,
\end{equation}
the effective-fluid energy conditions provide intrinsic bounds on $\chi$.
From the NEC,
\begin{equation}
(\rho+p)\left(1+\frac{\chi}{4\pi}\right)>0
\quad\Rightarrow\quad
\chi>-4\pi.
\end{equation}
From the DEC condition $\rho_{\mathrm{eff}}-p_{\mathrm{eff}}\ge 0$,
\begin{equation}
\rho_{\mathrm{eff}}-p_{\mathrm{eff}}=(\rho-p)\left(1+\frac{\chi}{2\pi}\right)\ge 0
\quad\Rightarrow\quad
\chi\ge -2\pi.
\end{equation}
For completeness, the SEC gives
\begin{equation}
\rho+3p+\frac{\chi}{\pi}p>0
\quad\Rightarrow\quad
\chi> -\pi\frac{\rho+3p}{p}\quad (p>0),
\end{equation}
which is model dependent through the local EoS. Therefore, the NEC/DEC bounds are theoretical consistency requirements of the model itself and do not rely on observational data.

\subsubsection{Regularity and Stability}
A second class of constraints follows from hydrostatic regularity of the modified TOV system. From Eq.~\eqref{dpdr}, we define
\begin{equation}
D \equiv \left(1+\frac{3\chi}{8\pi}\right)-\frac{\chi}{8\pi}\frac{1}{c_s^2},
\end{equation}
with the regularity requirement
\begin{equation}
D\neq 0.
\end{equation}
The corresponding critical sound speed is
\begin{equation}
c_{s,\star}^2=\frac{\chi}{8\pi+3\chi}.
\end{equation}
For realistic neutron star equations of state satisfying $c_s^2\to 0$ in the crust, the condition $D\neq 0$ excludes the $\chi>0$ branch, which generically develops a singularity in $dp/dr$. On the theoretically consistent neutron star branch (in particular $\chi\ge -2\pi$ from DEC), hydrostatic regularity then selects $\chi<0$ as the physically viable sector. This regularity/stability constraint is independent of observational data and follows directly from the structure of the modified TOV equations.

\subsubsection{Astrophysical Constraints}

Observational constraints are data driven. Viable configurations must satisfy a
maximum mass compatible with heavy-pulsar measurements
\cite{demorest/2010, antoniadis/2013, cromartie/2019, fonseca/2021}, radii
consistent with NICER and GW170817 inferences
\cite{riley/2019, miller/2019, riley/2021, miller/2021, abbott/2018}, and the
GW170817 tidal constraint \cite{abbott/2018}:
\begin{equation}
M_{\max} \gtrsim 2\,M_\odot,
\end{equation}
\begin{equation}
R_{1.4} \sim 11\text{-}14\ \mathrm{km},
\end{equation}
\begin{equation}
\Lambda_{1.4}\approx 190^{+390}_{-120}.
\end{equation}
In practice, multimessenger constraints restrict the magnitude of the coupling to $|\chi| \lesssim \mathcal{O}(0.1)$, although the precise bounds remain mildly dependent on the choice of microscopic equation of state.

\paragraph{Synthesis of Constraints}
The physically viable region is therefore given by the intersection of these domains: theoretical consistency (NEC/DEC), hydrostatic regularity, and multimessenger constraints. In practice, this intersection selects a narrow interval around General Relativity on the negative branch, approximately
\begin{equation}
-\mathcal{O}(0.1) \lesssim \chi < 0.
\end{equation}

\subsection{Relativistic Enthalpy Formulation}

For numerical integrations of neutron star structure it is advantageous
to rewrite the Tolman-Oppenheimer-Volkoff (TOV) equations using the
relativistic enthalpy as the independent variable \cite{lindblom/1992, lindblom/2010}. This formulation is
widely used in gravitational-wave analyses because it improves numerical
stability and works naturally with tabulated equations of state.

\subsubsection*{Relativistic Enthalpy}

For a barotropic equation of state $p=p(\rho)$ the relativistic enthalpy
is defined as
\begin{equation}
H(p) = \int_0^{p} \frac{dp'}{\rho(p') + p'}.
\end{equation}

Differentiating gives
\begin{equation}
\frac{dH}{dr} = \frac{1}{\rho+p}\frac{dp}{dr}.
\end{equation}

Substituting the TOV-like equations up to Eq.~\eqref{dpdr} into the definition of $dH/dr$ yields

\begin{equation}
\frac{dH}{dr}
=
-
\frac{\rho_{\rm eff}+p_{\rm eff}}{\rho+p}
\frac{1}{\left(1+\frac{3\chi}{8\pi}\right)-\frac{\chi}{8\pi}\frac{1}{c_s^2}}
\frac{m+4\pi r^3 p_{\rm eff}}{r(r-2m)}.
\end{equation}

\subsubsection*{Enthalpy Form of the Stellar Structure Equations}

It is convenient to use $H$ as the integration variable. Using the
identity
\begin{equation}
\frac{d}{dr} = \frac{dH}{dr}\frac{d}{dH},
\end{equation}
the stellar structure equations become

\begin{align}
\frac{dr}{dH} &= -
\frac{r(r-2m)}{m+4\pi r^3 p_{\rm eff}}
\frac{\rho+p}{\rho_{\rm eff}+p_{\rm eff}}
\Biggl[\left(1+\frac{3\chi}{8\pi}\right) \notag \\
&\quad-\frac{\chi}{8\pi}\frac{1}{c_s^2}\Biggr], \\
\nonumber
\\
\frac{dm}{dH} &= 4\pi r^2 \rho_{\rm eff}\frac{dr}{dH}.
\end{align}

The integration starts at the stellar center with
\begin{equation}
H=H_c,\quad r=0,\quad m=0,
\end{equation}
and proceeds outward until the surface is reached when $H=0$
(or equivalently $p=0$).

\subsubsection*{Tidal Love Number Equation}

The tidal deformability is determined by solving the differential
equation governing the metric perturbation function $y(r)$
introduced by Hinderer \cite{hinderer/2008, hinderer/2010, postnikov/2010}. In the standard radial formulation the
equation reads

\begin{equation}
\frac{dy}{dr} =
-\frac{y^2}{r}
-
\frac{y}{r}F(r)
-
rQ(r),
\end{equation}

where

\begin{equation}
F(r)=\frac{1+4\pi r^2(p_{\rm eff}-\rho_{\rm eff})}{1-2m/r},
\end{equation}

and

\begin{equation}
\begin{aligned}
Q(r)=&\frac{4\pi}{1-2m/r}
\Biggl(5\rho_{\rm eff}+9p_{\rm eff}+\frac{\rho_{\rm eff}+p_{\rm eff}}{c_{s,\rm eff}^2}\Biggr)\\
&-\frac{6}{r^2(1-2m/r)} -\frac{4\bigl(m+4\pi r^3 p_{\rm eff}\bigr)^2}{r^4(1-2m/r)^2},
\end{aligned}
\end{equation}
where the effective sound speed is defined by
\begin{equation}
c_{s,\rm eff}^2 \equiv \frac{dp_{\rm eff}}{d\rho_{\rm eff}}=
\frac{c_s^2+\frac{\chi}{8\pi}(3c_s^2-1)}{1+\frac{\chi}{8\pi}(3-c_s^2)}.
\end{equation}
This expression requires
\begin{equation}
1+\frac{\chi}{8\pi}(3-c_s^2)\neq 0.
\end{equation}
During the numerical integration of the tidal perturbation equations, a microscopic tolerance threshold is applied to this denominator to prevent floating-point singularities when evaluating the $Q(r)$ metric potential in the extreme low-density regime.

Tidal perturbation equations follow from the Einstein equations sourced by the effective fluid, the adiabatic sound speed entering the perturbation sector is the effective sound speed rather than the physical sound speed of the microscopic equation of state.

Using the enthalpy variable we obtain

\begin{equation}
\frac{dy}{dH} = \frac{dy}{dr}\frac{dr}{dH}.
\end{equation}

Thus the full system integrated numerically becomes

\begin{align}
\frac{dr}{dH} &= -
\frac{r(r-2m)}{m+4\pi r^3 p_{\rm eff}}
\frac{\rho+p}{\rho_{\rm eff}+p_{\rm eff}}
\Biggl[\left(1+\frac{3\chi}{8\pi}\right) \notag \\
&\quad-\frac{\chi}{8\pi}\frac{1}{c_s^2}\Biggr], \\\nonumber
\\
\frac{dm}{dH} &= 4\pi r^2 \rho_{\rm eff}\frac{dr}{dH}, \\\nonumber
\\
\frac{dy}{dH} &=
\left[-\frac{y^2}{r}-\frac{y}{r}F(r)-rQ(r)\right]\frac{dr}{dH}.
\end{align}

Because the theory can be written as Einstein gravity sourced by an effective fluid, the perturbation equations governing tidal deformations retain the same structure as in General Relativity, with the physical density and pressure replaced by the effective quantities.

\subsubsection*{Tidal Deformability}

After integrating to the stellar surface $r=R$ we obtain
$y_R=y(R)$. The dimensionless tidal Love number is \cite{binnington/2009, damour/2009}

\begin{equation}
k_2 = \frac{8C^5}{5}(1-2C)^2
\frac{2+2C(y_R-1)-y_R}{\Xi(C,y_R)},
\end{equation}

\begin{equation}
\begin{aligned}
\Xi(C,y_R)= &2C\left[6-3y_R+3C(5y_R-8)\right]\\
&+4C^3\Bigl[13-11y_R+C(3y_R-2)\\
&\qquad+2C^2(1+y_R)\Bigr]\\
&+3(1-2C)^2\left[2-y_R+2C(y_R-1)\right]\\
&\qquad\times\ln(1-2C).
\end{aligned}
\end{equation}

where $C=M/R$ is the compactness.

The dimensionless tidal deformability measured in
gravitational-wave observations is then

\begin{equation}
\Lambda = \frac{2}{3}k_2 C^{-5}.
\end{equation}

A noteworthy feature of the present model is that the field equations can be written exactly in the form of Einstein gravity sourced by an effective energy–momentum tensor. As a consequence, the linearized perturbation equations governing tidal deformations retain the same structure as in General Relativity. In contrast to other modified gravity theories, such as $f(R)$ gravity \cite{yazadjiev/2018}, no additional dynamical degrees of freedom appear in the perturbation sector. The modification of the tidal response therefore enters only through the effective thermodynamic variables and through the effective sound speed $c_{s,\mathrm{eff}}^2 = dp_{\mathrm{eff}}/d\rho_{\mathrm{eff}}$.

\section{Effective Equation of State and Stiffening Mechanism}
\label{eos}
In the numerical analysis, we employ realistic EoSs that support maximum masses near $2\,M_\odot$ and are broadly consistent with merger tidal constraints \cite{demorest/2010, antoniadis/2013, abbott/2017a, abbott/2018}. We therefore consider APR4, SLY, WFF1, and MPA1.

To understand the impact of the $f(R,T) = R + 2\chi T$ modification on stellar structure, we examine how the physical fluid variables map into their effective counterparts, which are the quantities that source the gravitational field. In this theory, the geometric modification acts as an additional contribution to the energy-momentum tensor, leading to an effective energy density $\rho_{\mathrm{eff}}$ and an effective pressure $p_{\mathrm{eff}}$, given in Eqs.~\eqref{eq:rhoeff} and \eqref{eq:peff}.

\begin{figure}[ht]
    \centering
    \includegraphics[width=0.95\columnwidth]{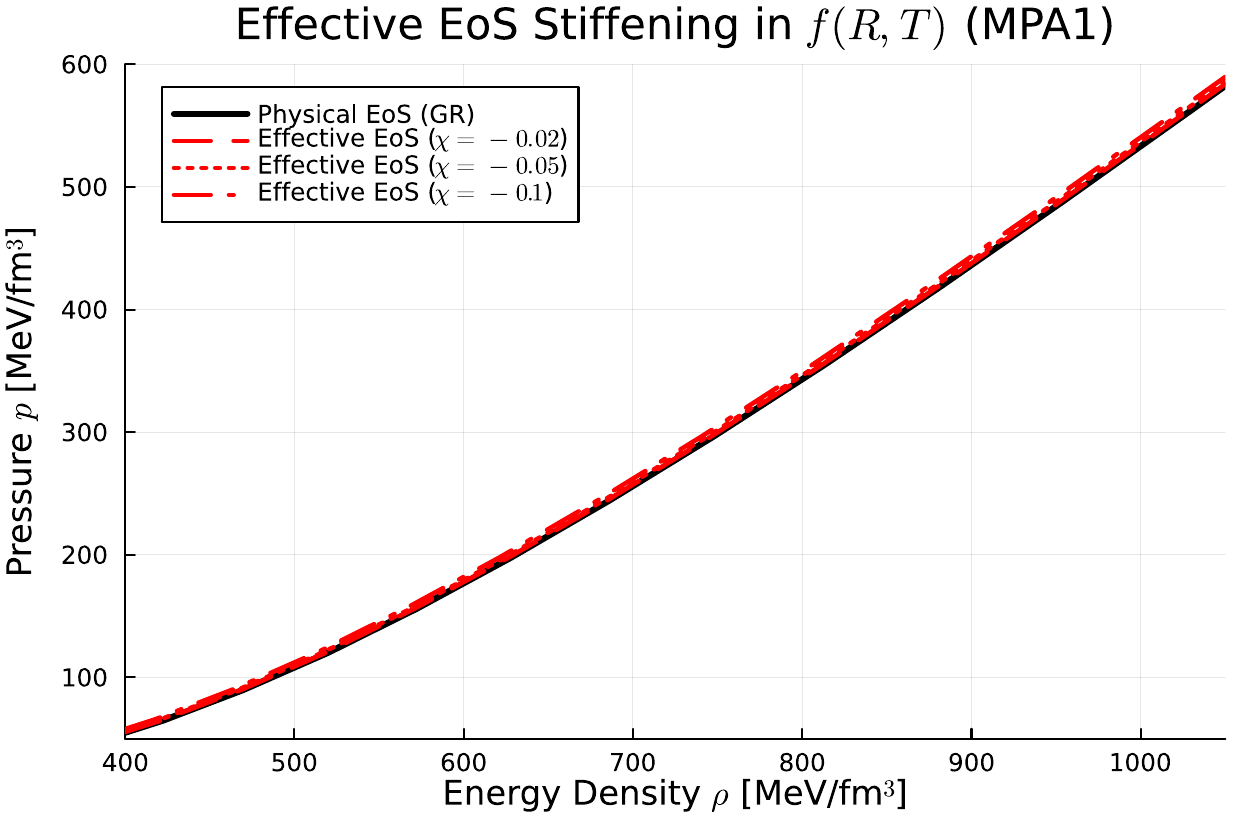}
    \caption{Effective EoS stiffening for the MPA1 model for different coupling strengths. All units are in $\text{MeV/fm}^3$.}
    \label{fig:stiffening_mpa1_mev}
\end{figure}

In Fig.~\ref{fig:stiffening_mpa1_mev}, we compare the physical EoS (GR baseline, $\chi=0$) with the effective EoS for the MPA1 model, in units of $\text{MeV/fm}^3$. For negative values of the coupling constant $\chi$, the effective pressure $p_{\mathrm{eff}}$ is systematically higher than the physical pressure $p$ at fixed energy density $\rho$. This shift constitutes an effective ``stiffening'' mechanism. Because the effective EoS is stiffer than the physical one, the configuration can withstand stronger gravitational compression before reaching the stability limit.

\begin{figure}[ht]
    \centering
    \includegraphics[width=0.95\columnwidth]{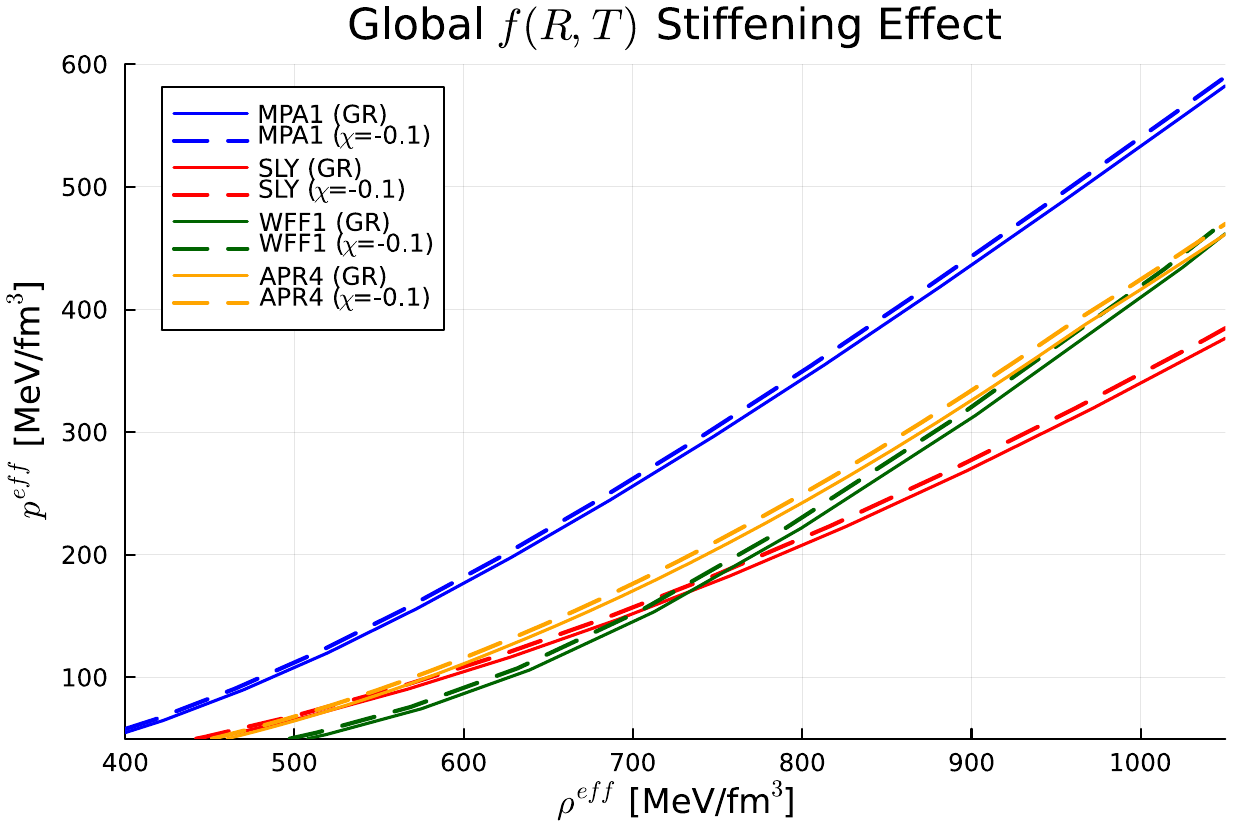}
    \caption{Global stiffening effect across various nuclear models (MPA1, SLY, WFF1, APR4), comparing GR to $\chi=-0.1$. All units are in $\text{MeV/fm}^3$.}
    \label{fig:stiffening_summary_mev}
\end{figure}

Figure~\ref{fig:stiffening_summary_mev} shows that this effect is model-independent. Across all considered equations of state, ranging from soft models like APR4 to stiffer models like MPA1, the presence of a negative $\chi$ consistently shifts the curve upward.

\section{Mass-Radius and Observational Constraints}
\label{sec:results}

In this section we present numerical solutions of the modified Tolman-Oppenheimer-Volkoff (TOV) and tidal equations in $f(R,T)=R+2\chi T$ gravity. We compare the predictions for APR4, SLY, WFF1, and MPA1 with multimessenger constraints from massive pulsars, NICER radius measurements (PSR J0030+0451 and PSR J0740+6620), and GW170817.

\subsection{Mass-Radius Relations and the Crustal Sound Speed Effect}

\begin{figure}[htpb]
    \centering
    \includegraphics[width=0.95\columnwidth]{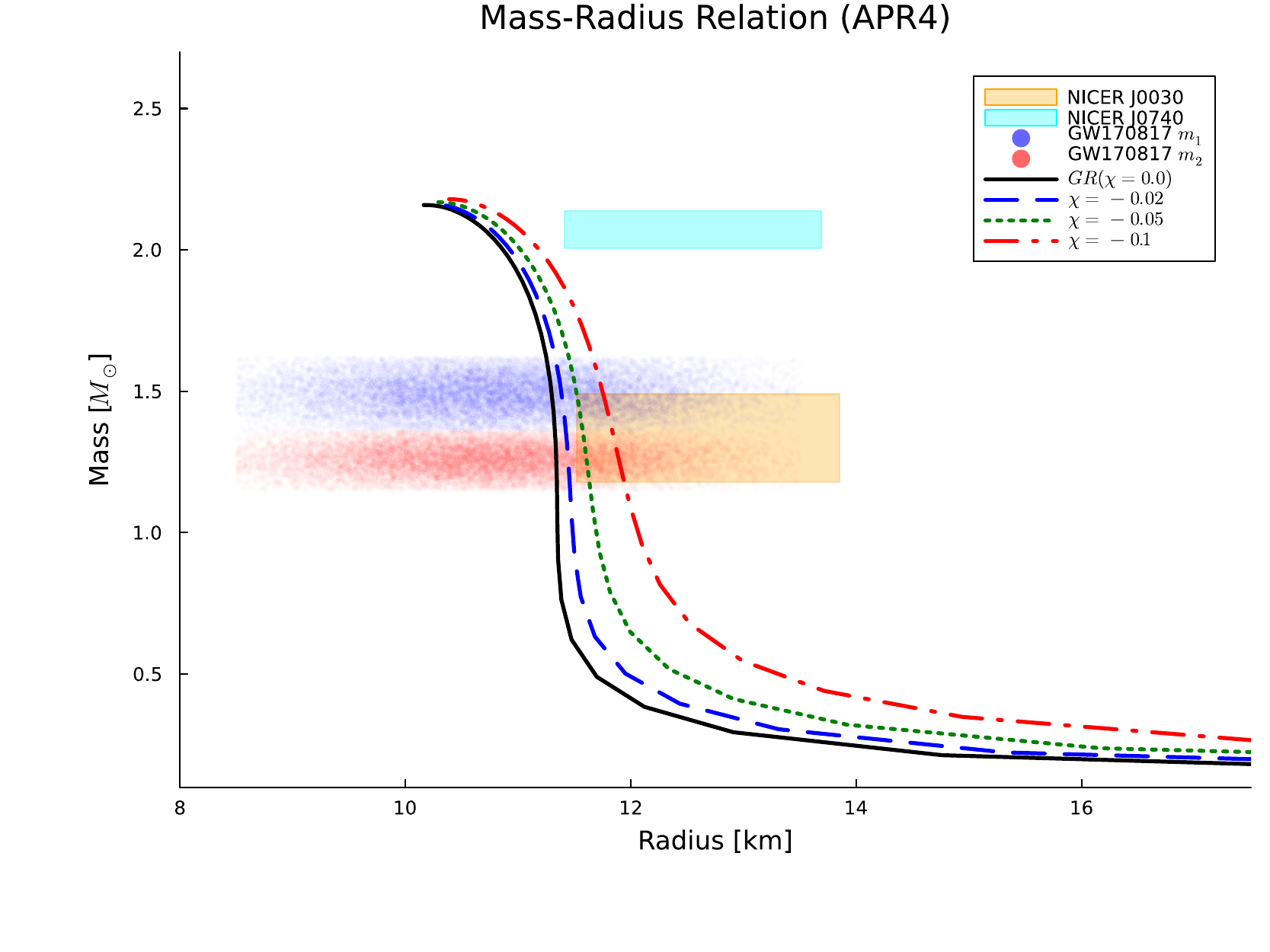}
    \caption{Mass-radius relation for the APR4 equation of state. The shaded rectangular regions denote the $1\sigma$ NICER constraints for PSR J0030+0451 (orange) and PSR J0740+6620 (cyan), while the background clouds represent the $90\%$ credible intervals for the primary and secondary masses from GW170817.}
    \label{fig:MR_APR4}
\end{figure}

\begin{figure}[htpb]
    \centering
    \includegraphics[width=0.95\columnwidth]{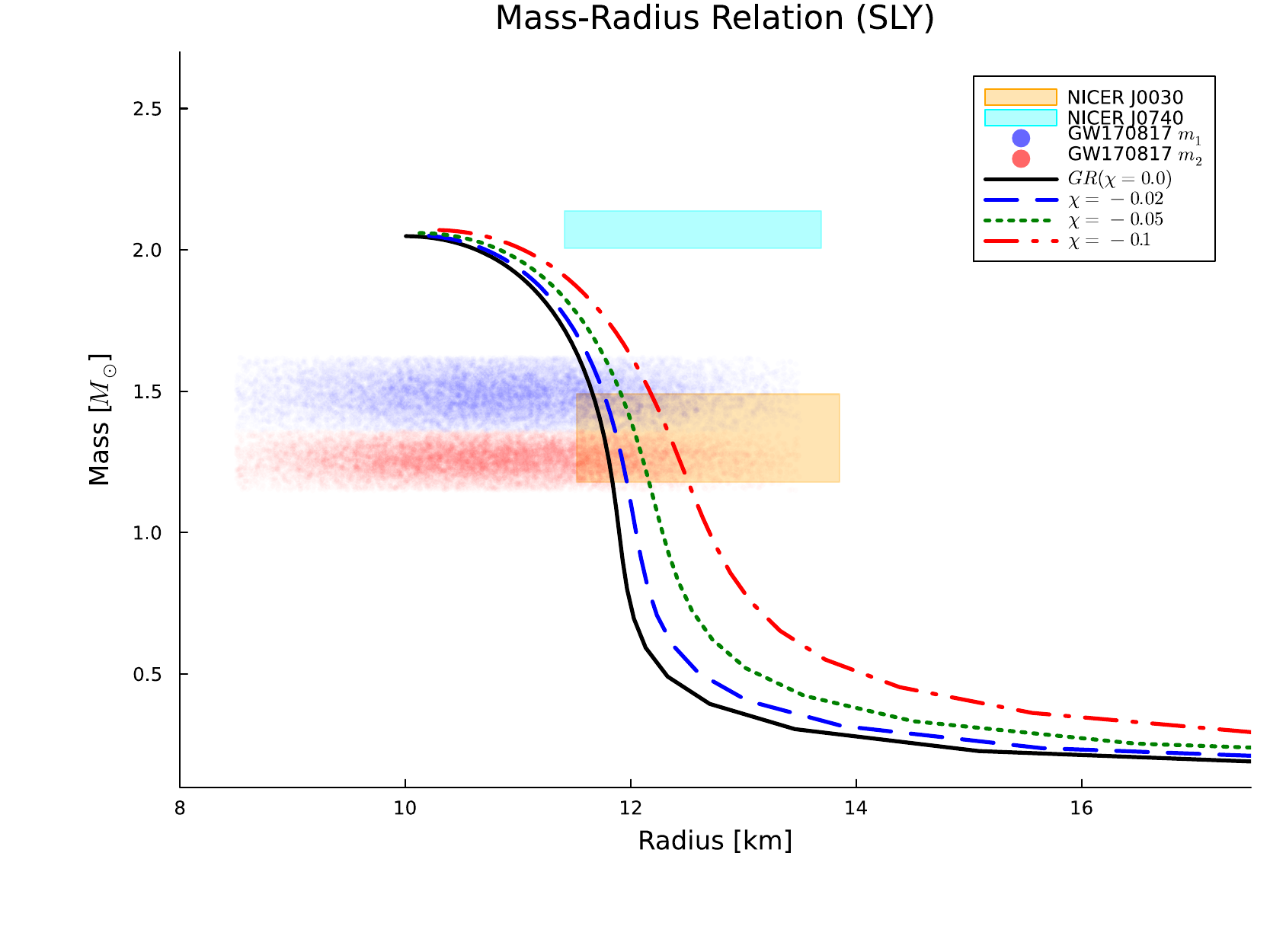}
    \caption{Mass-radius relation for the SLY equation of state. The plotting conventions are the same as in Fig.~\ref{fig:MR_APR4}.}
    \label{fig:MR_SLY}
\end{figure}

\begin{figure}[htpb]
    \centering
    \includegraphics[width=0.95\columnwidth]{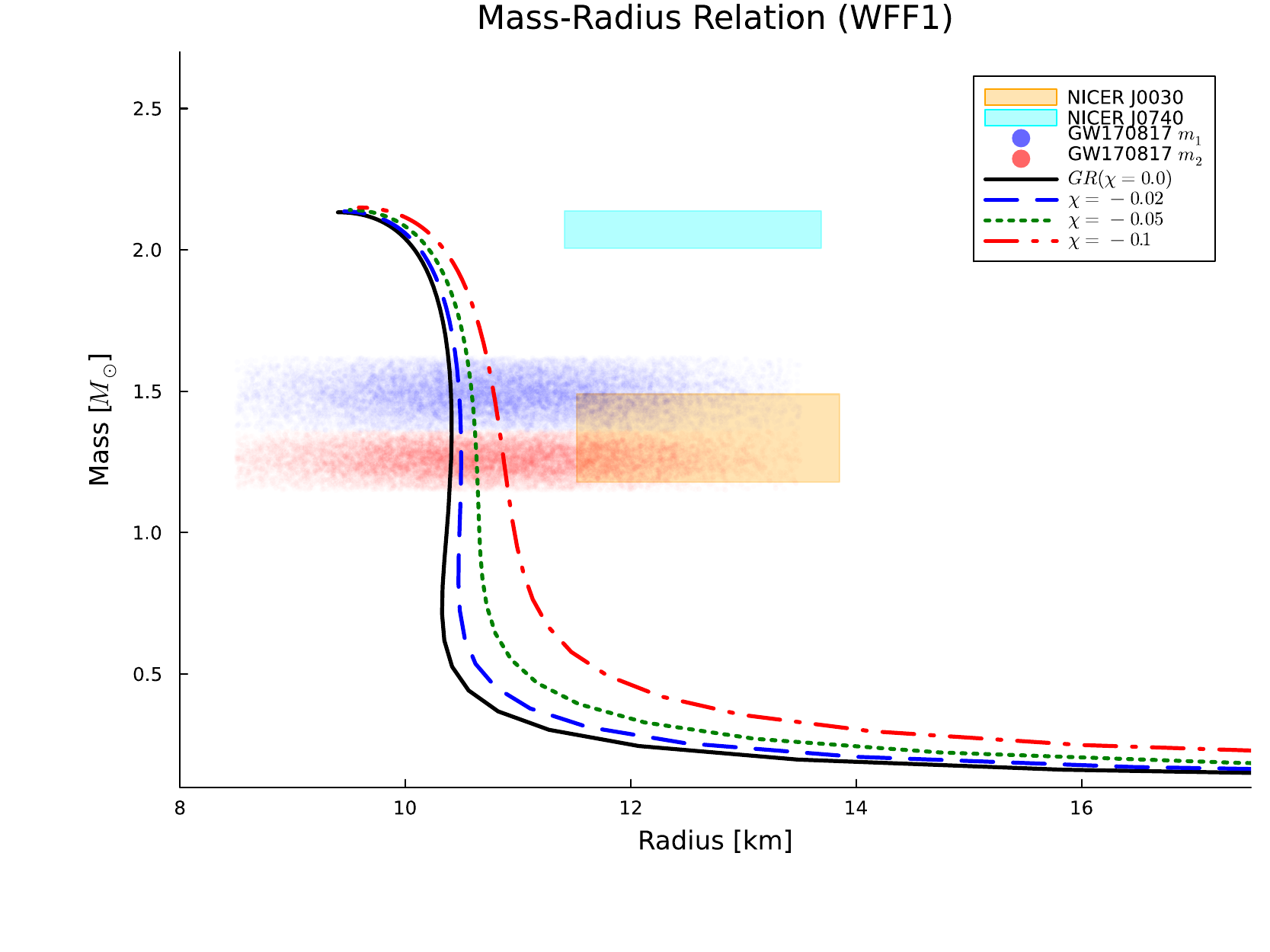}
    \caption{Mass-radius relation for the WFF1 equation of state. The plotting conventions are the same as in Fig.~\ref{fig:MR_APR4}.}
    \label{fig:MR_WFF1}
\end{figure}

\begin{figure}[htpb]
    \centering
    \includegraphics[width=0.95\columnwidth]{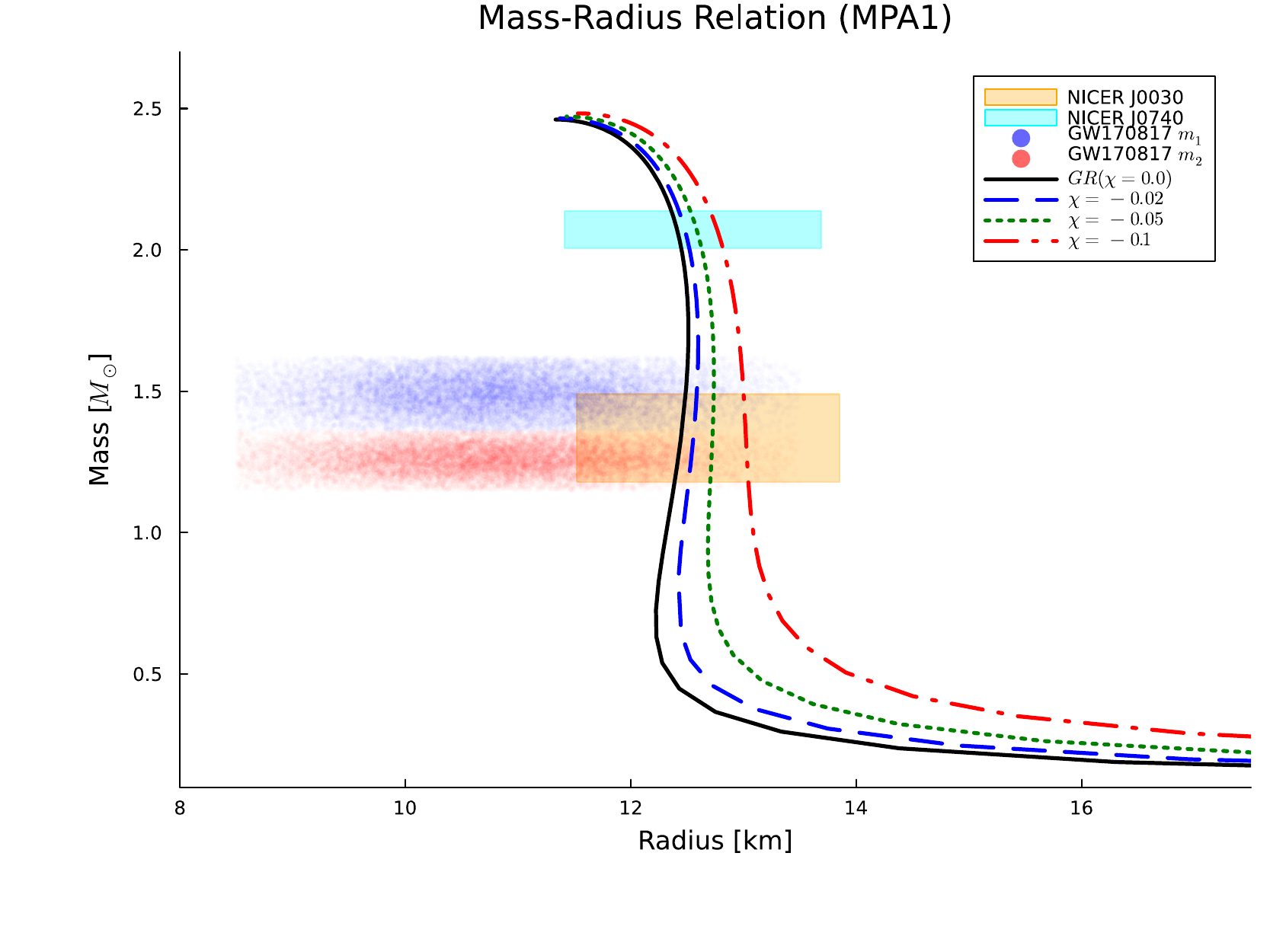}
    \caption{Mass-radius relation for the MPA1 equation of state. The stronger rightward shift for negative $\chi$ highlights the tension of very stiff models with multimessenger radius and tidal constraints.}
    \label{fig:MR_MPA1}
\end{figure}

Figures~\ref{fig:MR_APR4}, \ref{fig:MR_SLY}, \ref{fig:MR_WFF1}, and \ref{fig:MR_MPA1} display the mass-radius ($M-R$) relations for the selected EoSs. The black solid lines represent the General Relativity (GR) baseline ($\chi = 0.0$), while the dashed, dotted, and dash-dotted lines represent increasing strengths of the modified gravity coupling ($\chi = -0.02, -0.05$, and $-0.1$, respectively).

As $\chi$ becomes more negative, the effective stiffening of the EoS (as discussed in Section \ref{eos}) manifests as an outward shift in the stellar radius and a modest increase in the maximum mass. However, a critical phenomenon arises when applying extended gravity theories to realistic neutron star models that include a proper crust, as first demonstrated by Lobato et al. \cite{lobato/2020}.

A trend that is also evident in Figs.~\ref{fig:MR_APR4}-\ref{fig:MR_MPA1} is that the separation between curves with different $\chi$ is typically more pronounced in the low-mass region than near the maximum-mass branch. In other words, the coupling constant leaves a stronger imprint on less massive stars. Physically, these stars have lower compactness and a larger relative contribution from the outer layers, where the EoS is softer and the $1/c_s^2$ factor amplifies the geometric correction terms. For very massive stars, the stiff high-density core and stronger relativistic compactness partially reduce the relative impact of changing $\chi$ on global observables such as the radius.

In the modified hydrostatic equilibrium equations, the geometric correction factor exhibits a strict dependence on the inverse square of the fluid's sound speed, $c_s^2 = dp/d\rho$. Deep in the stellar core, the matter is highly incompressible, $c_s^2$ is large, and the modified gravity terms remain well-behaved. However, as the integration reaches the stellar crust, the density drops precipitously, and the EoS becomes extremely soft ($c_s^2 \to 0$). Lobato et al. showed that this $1/c_s^2$ dependence forces the $f(R,T)$ contributions to explode in the low-density regime, artificially inflating the crust.

Our current numerical integrations corroborate this finding. If the coupling constant $\chi$ is not strictly bounded (e.g., $|\chi| \le 0.02$) or if the sound speed is not regularized in the extreme vacuum, the stellar envelope undergoes unphysical runaway inflation, blowing the radius out to over 30 km. To ensure stable numerical integration through the outermost stellar layers and avoid solver stalling as $c_s^2 \to 0$, we introduce a minimum numerical floor for the sound speed strictly within the geometric correction factor. Specifically, the evaluation of the $f(R,T)$ modification assumes an effective minimum bound of $c_s^2 \ge 0.015$ in the extreme-vacuum limit. This regularization isolates the unphysical runaway inflation of the crust while preserving the exact hydrostatic dynamics throughout the dense stellar interior. Consequently, because realistic stellar structures demand a crust, the overall maximum mass increase is severely tempered compared to older, simplistic ``bare star'' models. The inclusion of the crust dictates that $f(R,T)$ gravity cannot arbitrarily raise the mass limit of soft EoS models without concurrently destabilizing the stellar surface.

Taken together, Figs.~\ref{fig:MR_APR4}-\ref{fig:MR_MPA1} reveal a clear hierarchy. SLY offers the best overall compromise with the current radius bands for mild negative couplings. APR4 remains comparatively viable but lies near the compact edge of the allowed region, whereas WFF1 is more marginal because it stays systematically on the low-radius side of the NICER-favored range. By contrast, MPA1 is the most strongly constrained: negative $\chi$ shifts it toward larger radii and, when combined with its large deformabilities, quickly drives it into tension with multimessenger bounds. This pattern supports the interpretation that modified-gravity stiffening competes with observational constraints from both sides of parameter space: overly compact branches remain too small, whereas very stiff branches become too extended and too deformable.

\subsection{Single Star Tidal Deformability}

\begin{figure}[htpb]
    \centering
    \includegraphics[width=0.95\columnwidth]{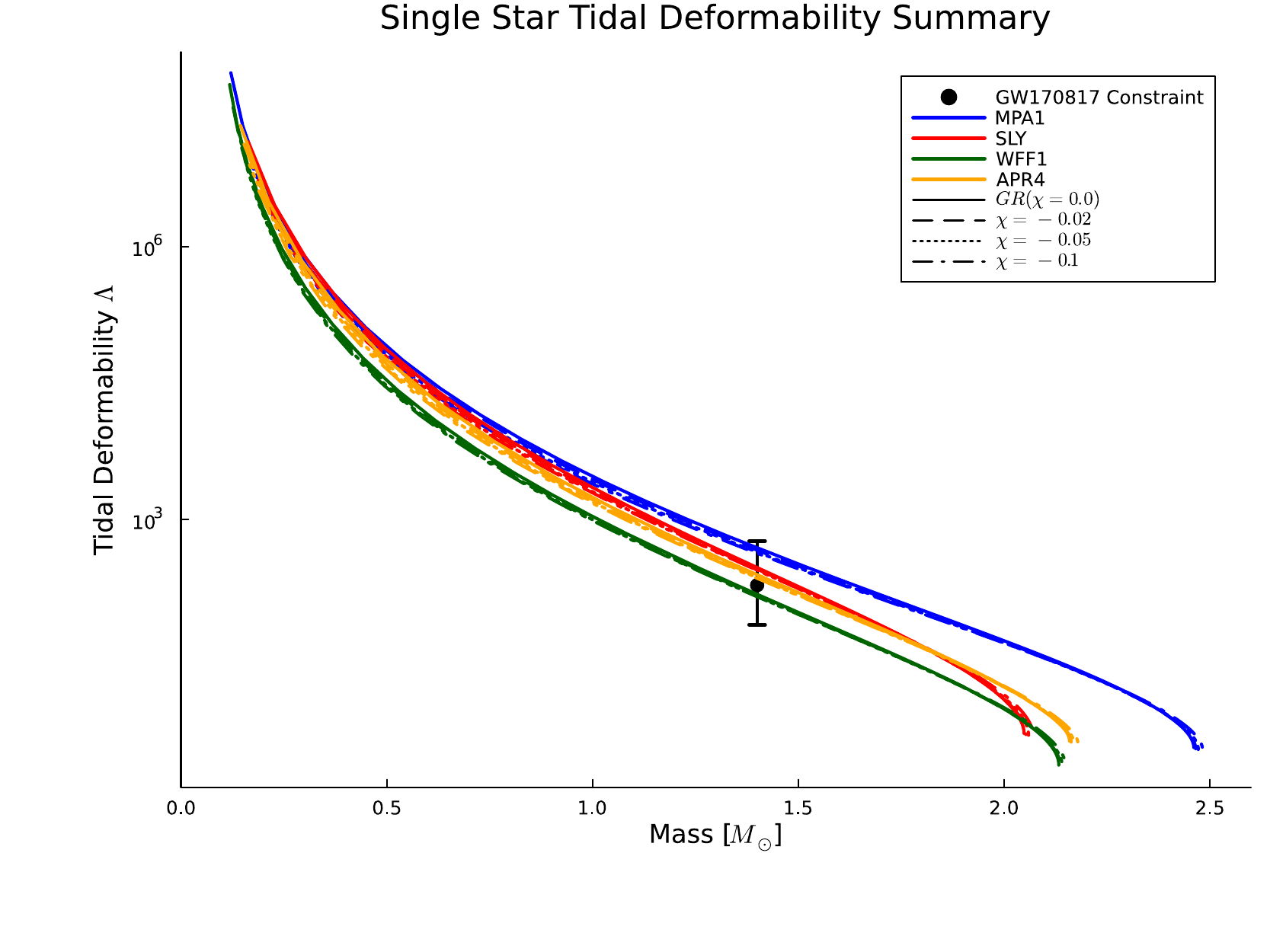}
    \caption{Dimensionless tidal deformability $\Lambda$ as a function of stellar mass for APR4, SLY, WFF1, and MPA1. The black marker with error bar corresponds to the GW170817-inspired reference constraint at $1.4\,M_\odot$.}
    \label{fig:Lambda_all}
\end{figure}

The internal stiffness of the star is directly probed by its tidal deformability, $\Lambda = (2/3)k_2 C^{-5}$, where $k_2$ is the Love number and $C = M/R$ is the compactness. Figure~\ref{fig:Lambda_all} summarizes $\Lambda(M)$ for APR4, SLY, WFF1, and MPA1.

Figure~\ref{fig:Lambda_all} shows that, for all EoSs, more negative $\chi$ shifts the $\Lambda(M)$ sequences upward, as expected from the larger radii and smaller compactness. The black marker with error bar at $1.4\,M_\odot$ denotes the GW170817 reference value $\Lambda_{1.4}=190^{+390}_{-120}$ and provides a direct visual discriminator among models.

The figure also reveals a clear hierarchy: MPA1 yields the largest deformabilities and WFF1 the smallest, while APR4 and SLY remain intermediate and partially overlap over much of the mass range. Consequently, the stiffer models come into stronger tension with tidal constraints as $\chi$ decreases, whereas the softer-to-intermediate cases remain compatible over a broader range of mild negative couplings.

\subsection{Binary Tidal Deformability}

\begin{figure}[htpb]
    \centering
    \includegraphics[width=0.95\columnwidth]{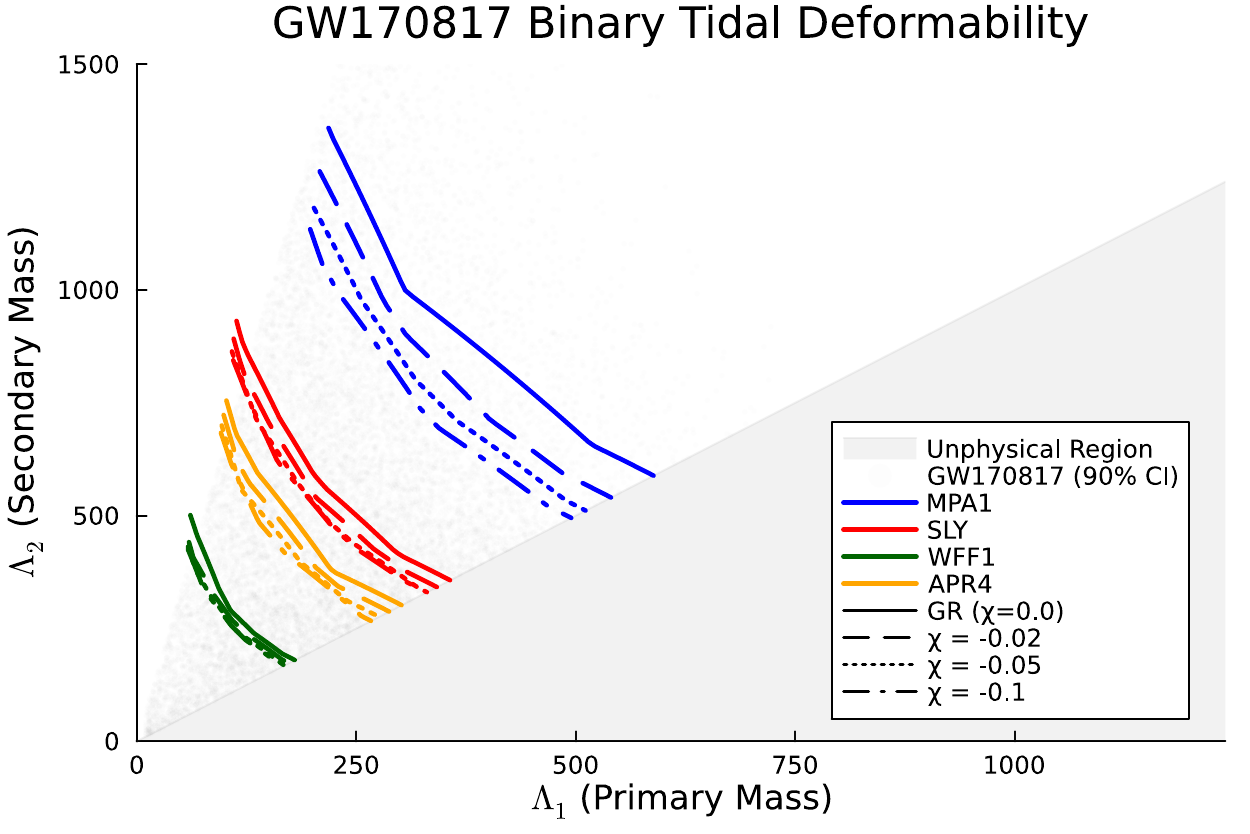}
    \caption{Binary tidal deformability ($\Lambda_1$ vs. $\Lambda_2$) for the studied EoSs. The colored lines trace the EoS trajectories mapped across the adopted GW170817 mass-ratio domain. The line styles indicate the strength of the $f(R,T)$ modification. The shaded gray region represents the unphysical domain ($\Lambda_2 < \Lambda_1$), while the scatter cloud serves as a proxy for the official LIGO/Virgo 90\% credible posterior.}
    \label{fig:Lambda_binary}
\end{figure}

To evaluate the theory against the LIGO/Virgo multimessenger data, we plot the binary tidal deformability ($\Lambda_1$ vs.~$\Lambda_2$) in Fig.~\ref{fig:Lambda_binary}. The gray cloud represents the synthetic 90\% credible posterior for GW170817, assuming a chirp mass of $\mathcal{M}=1.188\,M_\odot$ and a mass ratio $q\in[0.73,1.0]$.

The trajectories in the $\Lambda_1$-$\Lambda_2$ plane provide a stringent consistency test. APR4 and SLY pass through the densest part of the GW170817 posterior for mild couplings, WFF1 traces the lower-deformability edge of the allowed region, and MPA1 is shifted toward larger deformabilities. As $\chi$ becomes more negative (from solid to dash-dotted curves), the tracks move upward and to the right, progressively reducing the overlap with the observationally favored region.

This binary phase space encapsulates the main tension of $f(R,T)$ gravity in neutron stars: negative values of $\chi$ can help soft EoSs raise the maximum mass to the $2.0\,M_\odot$ level required by heavy pulsars, but overly negative values also drive the binary tidal deformability beyond the GW170817 credible region. Combined with the crustal sound-speed limitations highlighted by \cite{lobato/2020}, this leaves only a narrow viable interval for the coupling constant and confirms that departures from General Relativity in the dense-matter regime must remain small.

\section{Conclusions}
\label{sec:conclusions}

In this work, we formulated and tested a conservative $f(R,T)$ model in which the modified field equations are written as Einstein equations sourced by an effective fluid. This construction keeps the gravitational sector independent of the microscopic equation of state and avoids reconstructing the action from a chosen EoS. Within this framework, we derived the stellar-structure and tidal-perturbation equations in enthalpy form and solved them for realistic hadronic EoSs.

Our results show that negative values of the coupling constant $\chi$ effectively stiffen stellar matter, shifting the mass-radius and tidal-deformability curves upward. This effect is not uniform along the stellar sequence: it is strongest for low-mass stars, where lower compactness and a larger relative crust contribution amplify the modified-gravity corrections. Near the maximum-mass branch, the stiff high-density core reduces the relative impact of $\chi$ on global observables.

At the same time, the crustal regime is central to the viability bounds. Because the modified equilibrium equations contain terms proportional to $1/c_s^2$, the low-density region can trigger strong amplification and unphysical radius inflation if $|\chi|$ is too large. When this effect is treated consistently with realistic crusts, the allowed parameter space is significantly reduced. Confronting the model with multimessenger constraints from massive pulsars, NICER radii, and GW170817 tidal bounds, we find that only small departures from GR remain viable, with couplings close to zero and typically limited to mild negative values. Among the realistic EoSs considered here, SLY provides the best overall compromise with current multimessenger bounds for mild negative couplings; APR4 remains viable but tends to lie on the compact side of the allowed region; WFF1 is more marginal because its radii remain systematically small; and MPA1 is the most strongly constrained because the modified-gravity stiffening quickly pushes it toward both larger radii and larger tidal deformabilities.

A central conclusion is therefore methodological: realistic equations of state are indispensable for testing modified gravity with neutron stars. As emphasized by modern nuclear-matter compilations and constraints \cite{oertel/2017a}, only EoSs that consistently describe core and crust microphysics can provide reliable bounds on gravity parameters. Otherwise, simplified or overly idealized EoSs can either hide or exaggerate the gravity signal, worsening the gravity-microphysics degeneracy \cite{moraes/2016}. Our results reinforce that robust constraints on $f(R,T)$ models require realistic tabulated EoSs and full multimessenger consistency checks.

These conclusions are consistent with, and extend, the findings of our previous analysis in Ref.~\cite{lobato/2020}, where the crust/sound-speed pathology in realistic $f(R,T)$ neutron star models was first highlighted. Compared with the conservative study of Ref.~\cite{pretel/2021a}, the key methodological difference is that their analysis focused on hydrostatic equilibrium and asteroseismology using polytropic equations of state, which are useful for trend exploration but limited for direct multimessenger inference.

A key physical implication of the present work is that this difference is not merely methodological; it changes the stability conclusions decisively. Polytropic toy models without a realistic crust effectively remove the low-density regime in which $c_s^2$ becomes very small. In our formulation, the modified hydrostatic terms scale as $1/c_s^2$; therefore, once realistic crust physics is included and $c_s^2 \to 0$, the geometric corrections can grow catastrophically, driving unphysical envelope inflation and eliminating acceptable equilibrium solutions unless $|\chi|$ is minute. Accordingly, stability diagnostics inferred only from radial-oscillation spectra or binding-energy cusps computed with crust-free polytropes are not physically complete for realistic $f(R,T)$ neutron stars, because they miss this dominant crust-driven destabilization channel.

Our results show that, once realistic tabulated EoSs and crust physics are included, the viable coupling interval becomes narrower and the systematics clearer. The present study advances this line of work by recasting the theory in a conservative effective-tensor form and by combining mass-radius, single-star $\Lambda(M)$, and binary $(\Lambda_1,\Lambda_2)$ diagnostics in a unified multimessenger analysis. Overall, the evidence indicates that any viable $f(R,T)$ effect in neutron stars must be perturbative around General Relativity.

\bigskip

{\bf Acknowledgments}

RVL was supported by INCT-FNA (Instituto Nacional de Ci\^encia e Tecnologia, F\'{\i}sica Nuclear e Aplica\c c\~oes), research Project No.~464898/2014-5, and acknowledges support from CAPES/CNPq. GAC would like to thank CNPq, Fundação de Amparo a Pesquisa do Espírito Santo and Fundação Araucária for financial support under processes \#314121/2023-4, PROFIX 12/2024 and NAPI “Fenômenos extremos no Universo”.

\bibliographystyle{apsrev4-2}
\bibliography{bibliography}

\end{document}